# High-Field Electrical and Thermal Transport in Suspended Graphene

Vincent E. Dorgan,[1,2] Ashkan Behnam,[1,2] Hiram J. Conley,[3] Kirill I. Bolotin,[3] and Eric Pop[1,2,4]

[1]*Micro and Nanotechnology Lab, Univ. Illinois at Urbana-Champaign, IL 61801, USA*

[2]*Dept. Electrical & Computer Eng., Univ. Illinois at Urbana-Champaign, IL 61801, USA*

[3]*Dept. Physics & Astronomy, Vanderbilt University, Nashville, TN 37235, USA*

[4]*Beckman Institute, Univ. Illinois at Urbana-Champaign, IL 61801, USA*

Contact: epop@illinois.edu

**ABSTRACT:** We study the intrinsic transport properties of suspended graphene devices at high fields ($\geq$1 V/$\mu$m) and high temperatures ($\geq$1000 K). Across 15 samples, we find peak (average) saturation velocity of 3.6$\times$10$^7$ cm/s (1.7$\times$10$^7$ cm/s), and peak (average) thermal conductivity of 530 Wm$^{-1}$K$^{-1}$ (310 Wm$^{-1}$K$^{-1}$), at 1000 K. The saturation velocity is 2-4 times and the thermal conductivity 10-17 times greater than in silicon at such elevated temperatures. However, the thermal conductivity shows a steeper decrease at high temperature than in graphite, consistent with stronger effects of second-order three-phonon scattering. Our analysis of sample-to-sample variation suggests the behavior of "cleaner" devices most closely approaches the intrinsic high-field properties of graphene. This study reveals key features of charge and heat flow in graphene up to device breakdown, highlighting remaining unknowns under extreme operating conditions.

**KEYWORDS:** graphene, suspended, high-field, thermal conductivity, drift velocity, breakdown

Understanding and manipulating the intrinsic properties of materials is crucial both from a scientific point of view and for achieving practical applications. This challenge is particularly apparent in the case of atomically-thin materials like graphene, whose properties are strongly affected by interactions with adjacent substrates. For instance, the intrinsic mobility of electrons and holes in graphene is reduced by approximately a factor of ten,[1-4] and the thermal conductivity by about a factor of five when placed onto a typical substrate like $SiO_2$.[5-7] The former occurs due to scattering of charge carriers with substrate impurities and vibrational modes (remote phonons).[8] The latter is caused by the interaction of graphene phonons with the remote phonons of the substrate,[6] although subtle changes in the graphene phonon dispersion could also occur if the coupling with the substrate is very strong.[7]

Therefore, in order to understand the *intrinsic* electrical and thermal properties of graphene, it is necessary to study devices and samples freely suspended across microscale trenches.[1-3, 9-13] Nevertheless, such electrical transport studies have examined only low-field and low-temperature conditions. In addition, no data presently exist on the intrinsic (electrical *and* thermal) high-field behavior of graphene devices, which is essential for practical device operation, and where electrical and thermal transport are expected to be tightly coupled. By contrast, high-field measurements carried out on suspended carbon nanotubes (CNTs) had previously revealed a wealth of new physical phenomena, including negative differential conductance,[14, 15] thermal light emission,[16] and the presence of non-equilibrium optical phonons.[14, 17]

In this letter we examine the intrinsic transport properties of suspended graphene devices at high fields and high temperatures. This approach enables us to extract both the drift (saturation) velocity of charge carriers and the thermal conductivity of graphene up to higher temperatures than previously possible (>1000 K). Our systematic study includes experimental analysis of 15 samples combined with extensive simulations, including modeling of (coupled) electrical and thermal transport, and that of graphene-metal contact effects. We uncover the important role that thermally generated carriers play in such situations, and also discuss important high-field transport properties at the elevated temperatures up to device breakdown.

A schematic of a typical suspended graphene device is shown in Figure 1a. To assess the broadest range of samples, we fabricated such devices using both mechanically exfoliated graphene and graphene grown by chemical vapor deposition (CVD). Graphene was initially ex-foliated, or transferred in the case of CVD-graphene, onto ~300 nm of $SiO_2$ with a highly doped Si substrate (p-type, $5\times10^{-3}$ Ω·cm) as a back gate. The graphene was then patterned into rectangular devices using electron-beam (e-beam) lithography and an $O_2$ plasma etch. Next, we defined metal contacts consisting of 0.5-3 nm of Cr and 80 nm of Au. A previously established method[1] with some modifications was used to partially etch the supporting $SiO_2$ and suspend the graphene (see Supporting Information, Section A). In most cases approximately ~200 nm $SiO_2$ was etched under the graphene and partly under the contacts (see Figure 1), with ~100 nm remaining. A critical point dryer helped prevent the graphene from breaking or collapsing during the etching process. After fabrication we confirm suspension via scanning electron microscopy (SEM), as shown in Figure 1b-d. In order to avoid damaging the graphene from e-beam irradiation during SEM,[18] we only image "dummy" devices or use low acceleration voltages (~1 kV) to conduct SEM prior to making measurements. Some devices show a small amount of "wrinkling" (see, e.g., Figure 1d), possibly leading to some of the sample-to-sample variability described below.

Figure 2a shows a typical resistance (R) vs. back-gate voltage ($V_G$) measurement for a suspended exfoliated graphene device with length $L \approx 1.5$ μm and width $W \approx 850$ nm, in vacuum (~$10^{-5}$ torr) at room temperature. For all measurements, we limit the back-gate voltage to $|V_G| \leq$ 10 V to avoid collapsing the suspended channel.[19, 20] The as-fabricated devices do not immediately exhibit the "clean" electrical behavior one might associate with freely suspended graphene.[1-3] Although the channel is not in contact with the substrate, some residue from processing remains on the device, and this can be removed or minimized through a current annealing technique.[1, 3] In this process, we sweep the drain voltage ($V_D$) to increasingly higher values until the Dirac voltage ($V_0$) appears within the narrow usable $V_G$ window ($|V_G| \leq$ 10 V) and the electrical characteristics of the device stabilize (Figure 2a).[21]

Figure 2b displays the room temperature effective mobility ($\mu_0$) of the device from Figure 2a with a few estimates of the carrier density. There is some uncertainty in the mobility extraction because the carrier density is not well known at low-field, in part due to limited knowledge of the residual doping density $n^*$. Thus, Figure 2b displays the effective mobility at several residual doping levels[22] $n^* = 10^9$, $10^{11}$, and $2\times10^{11}$ cm$^{-2}$. Nevertheless, the estimated mobility is above 15,000 cm$^2$V$^{-1}$s$^{-1}$ at room temperature, consistent with previous work[3] which pointed out such values are limited by flexural phonons in suspended graphene at all but the lowest temperatures ($T \geq 10$ K).

Interestingly, we note that significantly above room temperature the carrier density in the suspended graphene channel becomes dominated by thermally generated carriers ($n_{th}$) and independent of gate voltage (Figure 2c). For example, at $V_{G0} = V_G - V_0 = 10$ V for a gate capacitance $C_G \approx 4$ nF/cm$^2$ (see Supporting Information, Section C) we estimate that the gate voltage induces $n_{cv} = C_G V_{G0}/q \approx 2.5 \times 10^{11}$ cm$^{-2}$ carriers; however, at 1200 K the total population of thermally generated[4] electrons and holes is comparable, $n_{th} = 2(\pi/6)(k_B T/\hbar v_F)^2 \approx 2.6 \times 10^{12}$ cm$^{-2}$. These trends are illustrated in Figure 2c at temperatures ranging from 300 K to 2000 K, calculated using a method described previously,[4] here using the $C_G$ determined by the series combination of the remaining SiO$_2$ and the gap resulting from the etched SiO$_2$.

We now turn to our high-field transport measurements of suspended graphene devices. Figure 3a displays the measured current density ($I/W$) as a function of average electric field along the channel ($F \approx V - IR_C)/L$, up to irreversible electrical breakdown of suspended exfoliated (red) and CVD-grown (blue) graphene devices, in vacuum (~$10^{-5}$ torr) at room temperature. Some devices show linearly increasing current at low fields followed by saturation-like behavior at high fields, while other devices show linear (and sometimes super-linear) current throughout. To understand this behavior, in Figure 3b we use our self-consistent electrical-thermal simulator of graphene (described previously[23-25] and available online[26]) to model $I/W$ vs. $F$ up to suspended graphene device breakdown. By varying the room-temperature low-field mobility from $\mu_0 =$ 2,500–25,000 cm$^2$V$^{-1}$s$^{-1}$ and incorporating the temperature dependence[4] of the mobility, $\mu(T)$, we are able to replicate the different types of curves observed experimentally (for additional details of the model see Supporting Information, Section C). We find that devices showing saturation-like behavior at high fields typically have a high $\mu_0$ and strong mobility dependence on temperature, (i.e., $\mu(T) \sim T^\beta$ where $\beta \approx 2.5$), being essentially "cleaner" and less disordered. Conversely, devices not showing saturation-like behavior have relatively low $\mu_0$ and a weaker temperature dependence ($\beta \approx 1.5$), which most likely corresponds to higher residual doping and disorder.[1, 8, 27] The super-linear current rise in such devices is due to the sharp increase in thermally generated carriers ($n_{th}$) as the device heats up.

Next, in Figure 4a we extract the carrier drift velocity ($v$) from our high-field transport data in Figure 3a, near the physical device breakdown. As discussed previously, the carrier density for our suspended devices has little to no gate dependence at high temperature reached at high fields, due to device self-heating. The maximum carrier drift velocity at the breakdown (BD) point is $v = I_{BD}/(qWn_{tot})$, where $n_{tot} = (n^{*2} + n_{th}^{2})^{1/2}$ is the carrier density[4] including residual doping ($n^{*} \sim 2 \times 10^{11}$ cm$^{-2}$) and thermal carrier generation, $n_{th}$. (The latter dominates at the elevated temperatures in the middle of the channel.) We note that the temperature profile, and thus the carrier density and drift velocity vary strongly along the channel near the BD point. However, because we cannot precisely model this profile for every device measured, we instead estimate the *average* drift velocity at breakdown, which is evaluated for an average carrier density along the channel, $\langle n_{tot} \rangle \approx 4 \times 10^{12}$ cm$^{-2}$ and $T_{avg} \approx 1200$ K, based on the thermal analysis discussed below.

Figure 4a displays the drift velocity at breakdown from all samples, arranged in increasing order of $(I_{BD}/W)/F_{BD}$, which our simulations (Figure 3b) suggest will rank them from most to least disordered. These data represent the *saturation* velocity in intrinsic graphene, as all measurements reached fields greater than 1 V/µm (see Figure 3a and footnote ref 28). The maximum values seen for samples #13-15 are very close to those predicted by a simple model[4] when transport is only limited by graphene optical phonons (OPs) with energy $\hbar\omega_{OP} = 160$ meV, i.e. $v_{sat} \approx 3.2 \times 10^{7}$ cm/s for the charge density and temperature estimated here ($4 \times 10^{12}$ cm$^{-2}$ and 1200 K, respectively). Similar saturation velocities have been predicted for clean, intrinsic graphene by extensive numerical simulations at comparable fields and carrier density.[29-31]

However, the average saturation velocity observed across our suspended samples is lower, $v = (1.7 +0.6/-0.3) \times 10^{7}$ cm/s, similar for exfoliated and CVD-grown graphene, at the average carrier densities and temperatures reached here. The average value remains a factor of two higher than the saturation velocity at elevated temperature in silicon ($v_{Si} = 8 \times 10^{6}$ cm/s at ~500 K),[32] but we suspect that variability between our samples is due to the presence of disorder and some impurities[33] which also affect the low-field mobility. In addition, depending on the level of strain built-in to these suspended samples (and how the strain evolves at high temperature), flexural phonons[3, 27] may also play a role in limiting high-field transport. It is apparent that future computational work remains needed to understand the details of high-field transport in graphene under a wide variety of temperatures and conditions, including ambipolar vs. unipolar transport, impurities and disorder.[28]

Next, we discuss the thermal analysis of our suspended graphene devices during high-field operation. While the breakdown temperature of graphene in air is relatively well-known as $T_{BD,air} \approx 600$ °C (based on thermogravimetric analysis[34, 35] and oxidation studies[36]), the breakdown tem-perature of graphene in probe station vacuum ($10^{-5}$ Torr) was not well understood before the start of this study. To estimate this, we compare similar devices taken up to electrical breakdown in air and in vacuum conditions. In both cases, we can assume heat transport is diffusive in our suspended devices at high temperature,[37] allowing us to write the heat diffusion equation:

$$\frac{d^2T}{dx^2} + \frac{P}{\kappa LWt} - \frac{2g}{\kappa t}\left(T - T_0\right) = 0 \, , \qquad (1)$$

where $\kappa$ is the thermal conductivity and $t = 0.34$ nm is the thickness of graphene, $g = 2.9 \times 10^4$ $Wm^{-2}K^{-1}$ is the thermal conductance per unit area between graphene and air,[12] and $g \approx 0$ in vacuum. Here the power dissipation is $P = I(V - IR_C)$ within the suspended graphene channel, and the temperature of the contacts at $x = \pm L/2$ is assumed constant, $T_0 \approx 300$ K (the small role of thermal contact resistance is discussed below). Assuming that $\kappa$ is a constant (average) along the graphene channel, we can compare breakdowns in air and vacuum and estimate the graphene device breakdown temperature in vacuum ($\sim 10^{-5}$ Torr) to be $T_{BD,vac} = 2230 + 630/-810$ K. (We note these upper and lower bound estimates are based on relatively extreme maximum/minimum choices, see Supporting Information, Section D.) The higher breakdown temperature in vacuum allows a higher power input in suspended graphene devices under vacuum conditions, consistent with previous studies of substrate-supported carbon nanotubes[38] and graphene nanoribbons.[39]

Having estimated a range for $T_{BD,vac}$, we turn to more detailed thermal modeling in order to extract $\kappa(T)$ from the electrical breakdown data. In general, we expect the thermal conductivity decreases with increasing temperature above 300 K, consistent with the case of carbon nanotubes, graphite and diamond.[5] Therefore, we write $\kappa = \kappa_0(T_0/T)^\gamma$ above room temperature and solve the one-dimensional heat diffusion equation, obtaining

$$T(x) = \left( T_0^{1-\gamma} + \frac{PL(1-\gamma)}{8\kappa_0 T_0^\gamma Wt}\left( 1 - \left(\frac{2x}{L}\right)^2 \right) \right)^{\frac{1}{1-\gamma}} , \qquad (2)$$

where $\kappa_0$ and $\gamma$ are fitting parameters, $\kappa_0$ being the thermal conductivity at $T_0 = 300$ K. (A similar analytic solution was previously proposed for suspended carbon nanotubes, albeit with a different functional form of the thermal conductivity.[40]) The breakdown temperature is maximum in the middle of the suspended graphene, at $x = 0$:

$$T_{BD,vac} = \left( T_0^{1-\gamma} + \frac{P_{BD}L(1-\gamma)}{8\kappa_0 T_0^{\gamma} W t} \right)^{\frac{1}{1-\gamma}}.$$

(3)

We note that SEM images (Supporting Information, Figure S3) show breakdown occurs in the center of the graphene channel, confirming the location of maximum temperature and good heat sinking at the metal contacts. For $T_{BD,vac} \approx 2230$ K we obtain $\gamma \approx 1.9$ and $\gamma \approx 1.7$ for our exfoliated and CVD-grown graphene samples, respectively.

Figure 4b shows the extracted thermal conductivity of each sample at $T = 1000$ K, for devices measured in vacuum. The lower bounds, circles, and upper bounds are based on $T_{BD,vac} = 2860, 2230$, and 1420 K respectively, the widest range of breakdown temperatures in vacuum estimated earlier. The average thermal conductivities at 1000 K of the exfoliated and CVD graphene samples are similar, $\kappa = (310 +200/-100)$ Wm$^{-1}$K$^{-1}$. Of this, the electronic contribution is expected to be <10%, based on a Wiedemann-Franz law estimate.[39] The result suggests that lattice phonons are almost entirely responsible for heat conduction in graphene even at elevated temperatures and under high current flow conditions. The average values of thermal conductivity found here are slightly lower than those of good-quality highly oriented pyrolytic graphite (534 Wm$^{-1}$K$^{-1}$ at $T = 1000$ K),[41] but the latter is consistent with the upper end of our estimates.

Figure 4c plots thermal conductivity as a function of temperature, showing that previously reported studies[9-13] (most near room temperature) fall on the same trend as this work at high temperature. However, our model of high-temperature thermal conductivity of suspended graphene suggests a steeper decrease ($\sim T^{1.7}$ weighed between exfoliated and CVD samples) than that of graphite ($\sim T^{1.1}$). The difference is likely due to the flexural phonons of isolated graphene, which could enable stronger second-order three-phonon[42, 43] scattering transitions ($\sim T^2$ scattering rate at high temperature) in addition to common first-order Umklapp phonon-phonon transitions ($\sim T$ scattering rate). Similar observations were made for silicon,[44] germanium[45] and carbon nanotubes[46, 47] at high temperatures, but not in isolated graphene until now.

We now comment more on the observed variability between samples, and on the role of graphene-metal contact resistance. First, even after high-temperature current annealing, polymer residue from processing may remain on the samples, increasing the scattering of both charge and heat carriers.[48] This sample-to-sample variation can contribute to the spread in extracted thermal conductivity and carrier velocity in Figures 4a-b. In this respect, it is likely that our samples yielding higher carrier velocity *and* higher thermal conductivity (e.g. sample #13 in Figures 4a-b) are the "cleanest" ones, most closely approaching the intrinsic limits of transport in suspended graphene. Second, some edge damage occasionally seen after high-current annealing affects our ability to accurately determine the sample width, $W$. The value used for $W$ influences all our calculations, and its uncertainty is incorporated in the error bars in Figure 4. (The case of an extreme $W$ reduction leading to a suspended graphene nanoconstriction is shown in the Supporting Information, Section E. Such devices were not used for extracting the transport data in the main text.) Third, several of the CVD graphene samples in Figure 4 had $W < 200$ nm, and edge scattering effects are known to limit transport in narrow ribbons.[39, 49] However, we saw no obvious dependence of thermal conductivity, carrier velocity, or breakdown current density ($I_{BD}/W$) on sample width when comparing "wide" and "narrow" devices. We thus expect that variation among samples due to edge scattering is smaller than other sources of variation.

Before concluding, we return to thermal contact resistance and estimate the temperature rise at the contacts[25] due to Joule heating during high-field current flow. The thermal resistance for heat flow from the suspended graphene channel into the metal contacts is given by[50]

$$R_{C,th} = \frac{1}{hL_T W_C} \coth\left(\frac{L_C}{L_T}\right), \qquad (4)$$

where $h$ is the thermal interface conductance per unit area for heat flow from the graphene into the $SiO_2$ or Au, $W_C$ is the width of the graphene under the contact, and $L_C$ is the metal contact length. The thermal transfer length $L_T = (\kappa t/h)^{1/2}$ corresponds to the distance over which the temperature drops by $1/e$ within the contact.[25] Typical contact lengths are on the order of microns while $L_T \approx 50$ nm, thus we have $L_C \gg L_T$ and we can simplify $R_{C,th} \approx (hL_T W_C)^{-1} \approx (W_C)^{-1}(h\kappa t)^{-1/2}$. Heat dissipation at the contacts consists of parallel paths to the underlying $SiO_2$ substrate, and top metal contact with $h_{g-ox} \approx 10^8$ Wm$^{-2}$K$^{-1}$ and $h_{g-Au} \approx 4 \times 10^7$ Wm$^{-2}$K$^{-1}$ (refs 51, 52), then the total thermal resistance for one contact is $R_{C,th} = (R_{C,ox}^{-1} + R_{C,Au}^{-1})^{-1}$. The temperature rise at the contacts is estimated as $\Delta T_C = T_C - T_0 = R_{C,th}P_{BD}/2$ (where $P_{BD}$ is the input electrical power at the breakdown point), which is only 10s of Kelvin. Including thermal contact resistance for the analysis when estimating $T_{BD,vac}$ above would change the extracted values by less than 4%.

In summary, we fabricated suspended graphene devices and carefully analyzed their high-field electrical and thermal transport. The electrical transport is entirely dominated by thermally-generated carriers at high temperatures (>1000 K), with little or no control from the substrate "gate" underneath such devices. The maximum saturation velocity recorded is >$3\times10^7$ cm/s, consistent with theoretical predictions for intrinsic transport limited only by graphene optical phonons. However, average saturation velocities are lower (although remaining a factor of two greater than in silicon at these temperatures), due to sample-to-sample variation. We estimated the breakdown temperature of graphene in $10^{-5}$ Torr vacuum, ~2230 K, which combined with our models yields an average thermal conductivity of ~310 Wm$^{-1}$K$^{-1}$ at 1000 K for both exfoliated and CVD-grown graphene. The models show a thermal conductivity dependence as ~$T^{-1.7}$ above room temperature, a steeper drop-off than that of graphite, suggesting stronger effects of second-order three-phonon scattering. Our study also highlights remaining unknowns, requiring future efforts on electrical and thermal transport at high field and high temperature in graphene.

## ■ ASSOCIATED CONTENT

**Supporting Information**

Details of fabrication and suspension process; method for estimating electrical contact resistance; additional details of suspended graphene device modeling; additional SEM images and details of breakdown in vacuum, air, and $O_2$ environments; suspended graphene nanoconstriction with high on/off current ratio. Material is available free of charge via the Internet at http://pubs.acs.org.

## ■ ACKNOWLEDGMENT

This work has been supported in part by the Nanotechnology Research Initiative (NRI), the Office of Naval Research Young Investigator Award (ONR-YIP N00014-10-1-0853) and the National Science Foundation (NSF CAREER ECCS 0954423). V.E.D. acknowledges support from the NSF Graduate Research Fellowship.

## ■ REFERENCES

21. We note that increases or decreases in resistance that do not always correspond to changes in $V_0$ are often seen during the annealing process. Although we expect the removal of impurities to affect the graphene resistivity, we believe these shifts are often caused by a change in the contact resistance. For cases when the device resistance increases substantially we find that the graphene channel has undergone partial breakdown (see Supporting Information, Sections D and E). We took care to avoid using such devices when extracting transport properties.

28. Some theoretical studies have predicted negative differential drift velocity at high fields, i.e. drift velocity which does not saturate but instead reaches a peak and then slightly decreases at higher fields (ref 29). This theoretical suggestion can neither be confirmed nor ruled out here, within the error bars in Figure 4a. We also point out that the transport regime observed here is not identical to our previous work on $SiO_2$-supported graphene (refs 4, 23). Here, the high-field measurement can only be carried out in an ambipolar, thermally intrinsic regime ($n \approx p \approx n_{th}/2$), whereas previous work for substrate-supported devices focused on unipolar (e.g. $n \gg p$) transport.

■ **FIGURES**

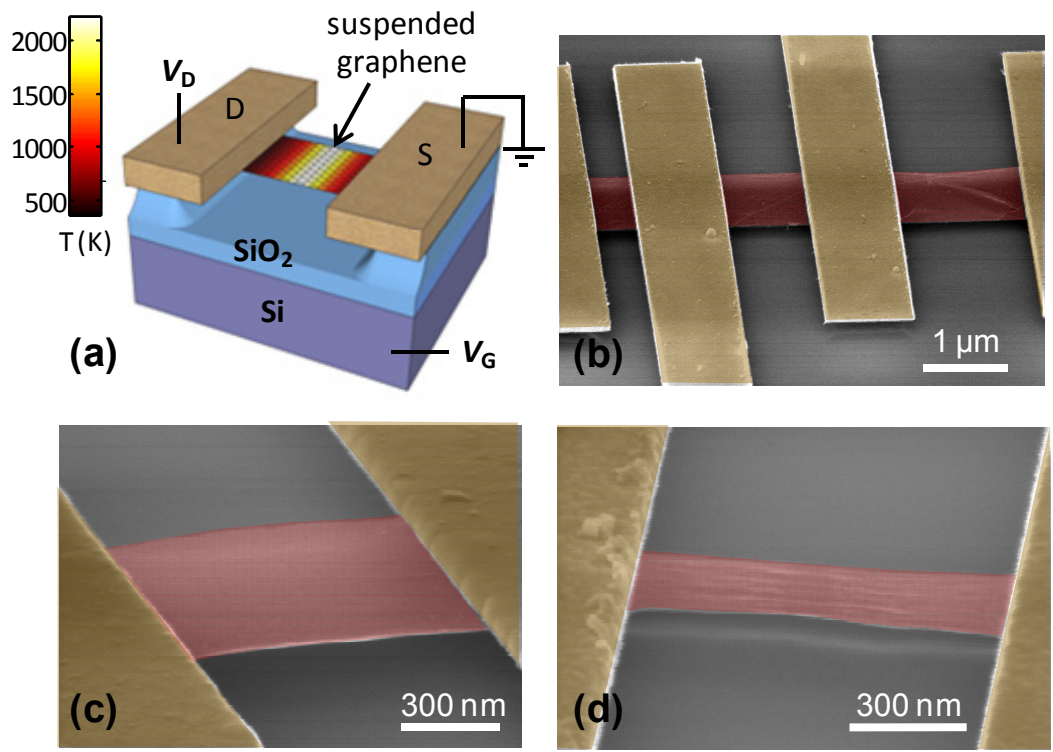

**Figure 1.** (a) Schematic of suspended graphene device. The color scale indicates the temperature of a suspended device during high-field current flow in vacuum (here calculated for the sample corresponding to Figure 2a-b, with applied power $P$ = 1.2 mW). Scanning electron microscopy (SEM) images of: (b) suspended graphene grown by chemical vapor deposition (CVD), and (c-d) suspended exfoliated graphene samples. All SEM images taken at a 70° tilt with respect to the substrate. The initial $SiO_2$ thickness was 300 nm, of which approximately 100 nm are left after under-etching.

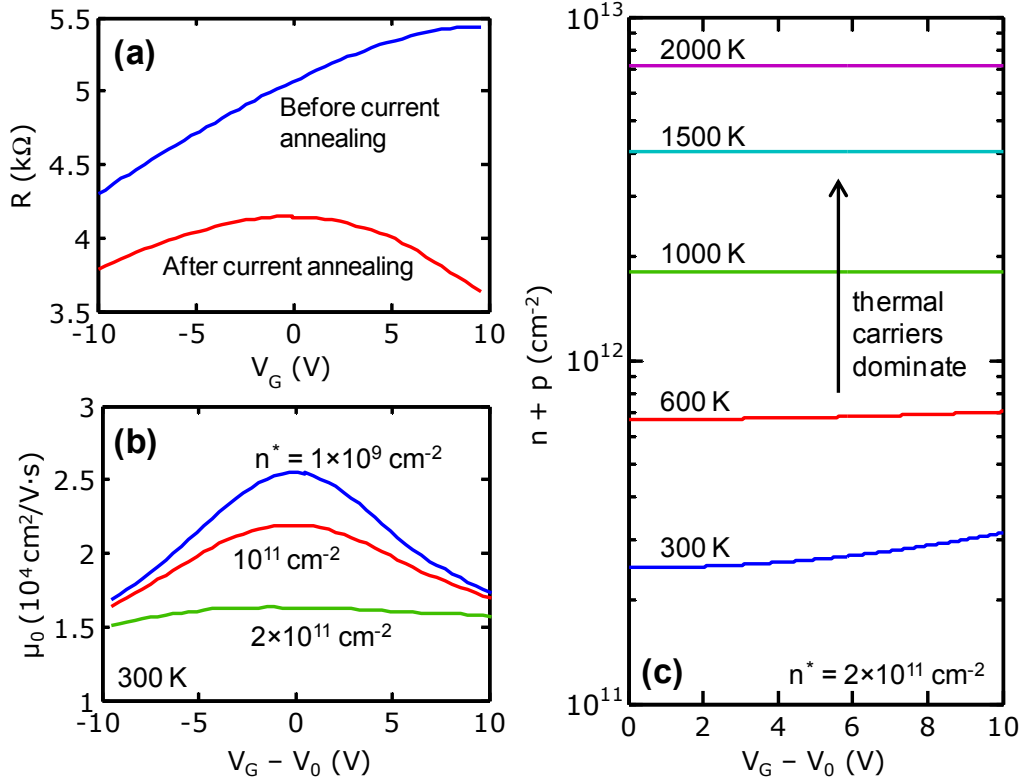

**Figure 2.** (a) Measured resistance vs. gate voltage $V_G$ at room temperature, before and after current annealing for a suspended exfoliated graphene device ($L$ = 1.5 μm, $W$ = 0.85 μm, $V_{DS}$ = 50 mV). (b) Effective mobility for the data shown in (a), assuming three different carrier densities as labeled. The contact resistance was estimated using the transfer length method (TLM) for devices of different channel lengths (see Supporting Information, Section B). (c) Calculated total carrier density ($n + p$) vs. gate voltage at increasing temperatures. In such suspended devices the carrier density becomes only a function of temperature (due to thermal carrier generation, $n_{th}$)[4] and independent of gate voltage at temperatures >600 K. This corresponds to all high-field transport cases studied in this work (Figures 3 and 4).

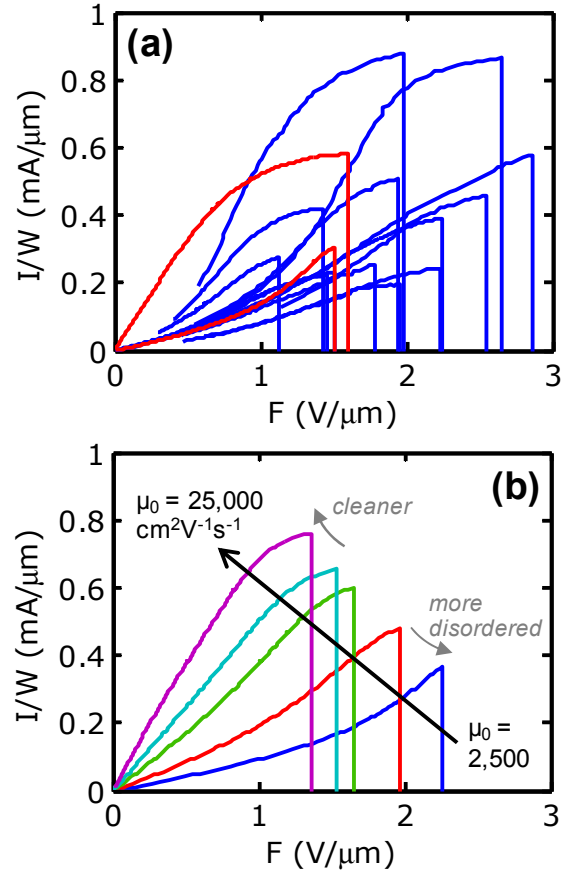

**Figure 3.** (a) Measured current density ($I/W$) vs. average electric field up to breakdown in vacuum of suspended graphene devices ($V_G = 0$). Exfoliated graphene (red) and CVD-grown graphene (blue) devices. The average electric field is $F = (V_{DS} - IR_C)/L$, accounting for the electrical contact resistance $R_C$ (see Supporting Information, Section B). (b) Simulated $I/W$ vs. $F$ with varying low-field mobility ($\mu_0$) from 2,500–25,000 $cm^2V^{-1}s^{-1}$. The simulations are based on our electro-thermal self-consistent simulator[23-25] (available online[26]), adapted here for suspended graphene. Suspended devices which reach higher, saturating current and break down at lower voltage are expected to be representative of cleaner, less disordered samples.

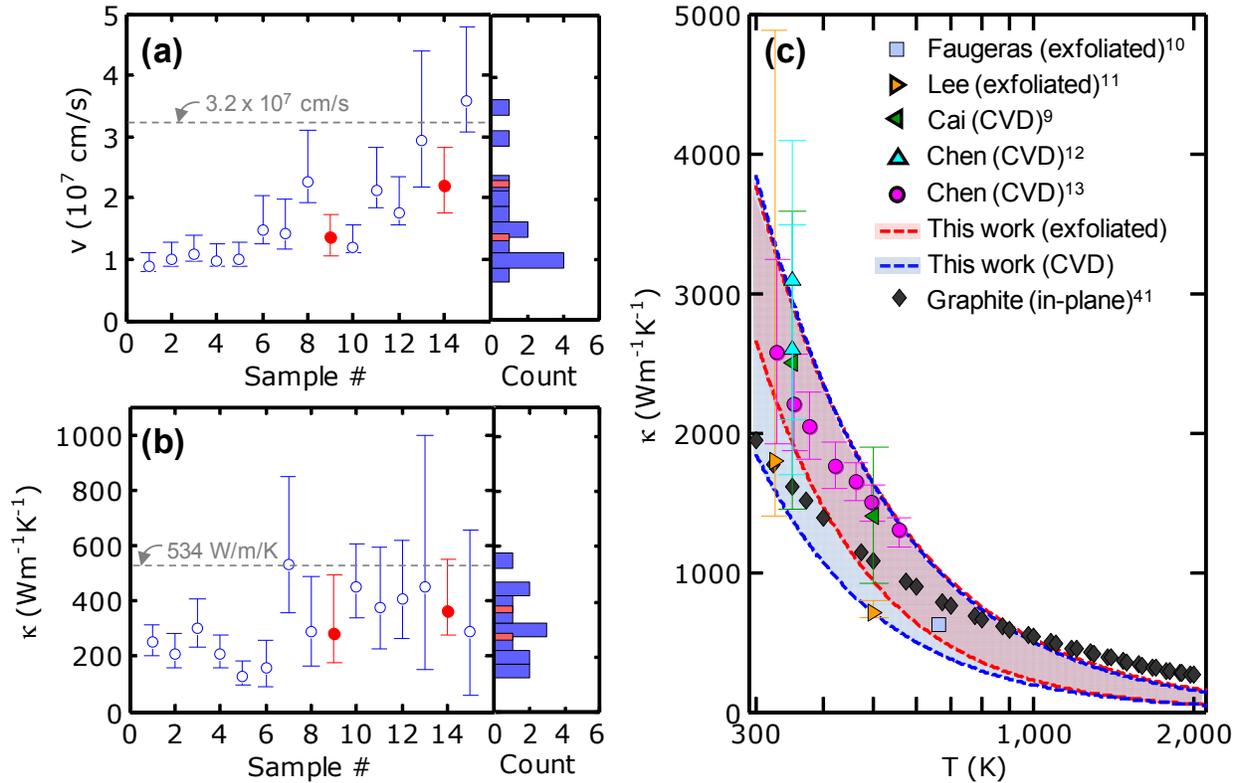

**Figure 4.** (a) Charge carrier saturation velocity at high temperature ($T_{avg} \approx 1200$ K) along the suspended graphene channel. (b) Corresponding thermal conductivity of the same suspended samples at ~1000 K. Exfoliated (red) and CVD (blue) graphene devices. Samples are ordered by increasing ($I_{BD}/W$)/$F_{BD}$, representative of increasingly "cleaner" devices as shown in Figure 3b. Lower bounds, symbols, upper bounds correspond to models with breakdown temperatures of 2860, 2230, 1420 K respectively (in $10^{-5}$ Torr vacuum). Some uncertainty also comes from imprecise knowledge of the device width $W$ (see Supporting Information). (c) Suspended graphene thermal conductivity above room temperature estimated from this work (lines) and that of previous studies (symbols). Shaded regions represent the average ranges of values for exfoliated (red) and CVD (blue) graphene from this work. The weighted average thermal conductivity for our samples is ~2500 Wm$^{-1}$K$^{-1}$ at room temperature and ~310 Wm$^{-1}$K$^{-1}$ at 1000 K, with a steeper drop-off than graphite, attributed to second-order three-phonon scattering (see text).

# <u>Supplementary Information</u>

## High-Field Electrical and Thermal Transport in Suspended Graphene

Vincent E. Dorgan,[1,2] Ashkan Behnam,[1,2] Hiram Conley,[3] Kirill I. Bolotin,[3] and Eric Pop[1,2,4]

[1]*Micro and Nanotechnology Lab, Univ. Illinois at Urbana-Champaign, IL 61801, USA*

[2]*Dept. Electrical & Computer Eng., Univ. Illinois at Urbana-Champaign, IL 61801, USA*

[3]*Dept. Physics & Astronomy, Vanderbilt University, Nashville, TN 37235, USA*

[4]*Beckman Institute, Univ. Illinois at Urbana-Champaign, IL 61801, USA*

## A.  Graphene device fabrication and suspension

We fabricated suspended devices by two methods, one by mechanically exfoliating graphene from natural graphite, the other by chemical vapor deposition (CVD) growth on Cu substrates. With the standard "tape method," graphene is mechanically exfoliated onto a substrate of ~300 nm of $SiO_2$ with a highly doped Si substrate (*p*-type, $5\times10^{-3}$ $\Omega$-cm). The tape residue is then cleaned off by annealing at 400 °C for 120 min with a flow of Ar/H (500/500 sccm) at atmospheric pressure. Monolayer graphene flakes are then identified with an optical microscope and confirmed via Raman spectroscopy (Figure S1a).[1]

Graphene growth by CVD is performed by flowing $CH_4$ and Ar gases at 1000 °C and 0.5 Torr chamber pressure, which results primarily in monolayer graphene growth on both sides of the Cu foil.[2] One graphene side is protected with a ~250 nm thick layer of polymethyl methacrylate (PMMA) while the other is removed with a 20 sccm $O_2$ plasma reactive ion etch (RIE) for 20 seconds. The Cu foil is then etched overnight in aqueous $FeCl_3$, leaving the graphene supported by the PMMA floating on the surface of the solution. The PMMA + graphene bilayer film is transferred via a glass slide to a HCl bath and then to two separate deionized water baths. Next, the film is transferred to the $SiO_2$ (~300 nm) on Si substrate (*p*-type, $5\times10^{-3}$ $\Omega$-cm) and left for a few hours to dry. The PMMA is removed using a 1:1 mixture of methylene chloride and methanol, followed by a one hour Ar/$H_2$ anneal at 400 °C to remove PMMA and other organic residue.

The following fabrication steps are performed for both the exfoliated graphene and CVD graphene devices. We pattern a rectangular graphene channel using e-beam lithography and an $O_2$ plasma etch. Another e-beam lithography step is used to define the electrodes, which consist of 0.5-3 nm of Cr and 80 nm of Au. The sample is annealed again in Ar/H at 400 °C in order to help remove the polymer residue leftover from the fabrication process.

The suspension of the graphene sheet is accomplished by etching away ~200 nm of the underlying $SiO_2$.[3] The sample is placed in 50:1 BOE for 18 min followed by a deionized (DI) water bath for 5 min. Isopropyl alcohol (IPA) is squirted into the water bath while the water is poured out so that the sample always remains in liquid. After all the water has been poured out and only IPA remains, the sample is put into a critical point dryer (CPD). Following the CPD process we confirm suspension using SEM (Figures 1b-d) or AFM (Figures S1b,c). We note that the CVD graphene samples underwent a vacuum anneal at 200 °C after the suspension process. Typically, device performance improved after this anneal but we also noticed several devices would break. Due to the fragility and relatively limited number of exfoliated graphene devices, we did not perform this additional annealing step with the exfoliated graphene samples.

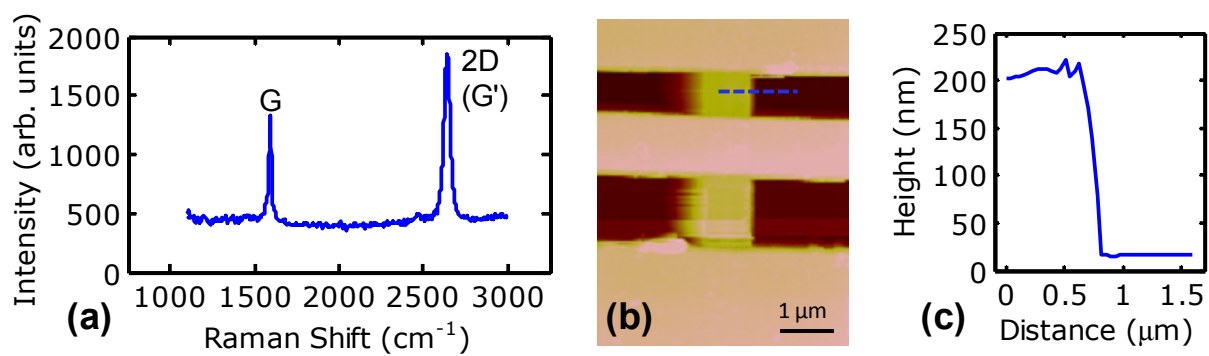

**Figure S1.** (a) Raman spectrum showing the G and 2D (also called G') bands of monolayer graphene. (b) AFM image of suspended exfoliated graphene where the dashed blue line corresponds to the (c) height vs. distance trace. The initial $SiO_2$ thickness is 300 nm, of which approximately 200 nm is etched during the suspension process.

## B.  Electrical contact resistance

Contact resistance of the CVD graphene devices is determined using the transfer length method (TLM). In Figure S2 we plot $R \cdot W$ versus $L$ and extract $R_C W \approx 1100$ Ω-um from the linear fit (dashed line). We use the resistance value at breakdown for extracting contact resistance. The spread in $R \cdot W$ values that results in a poor linear fit in Figure S2 may be associated with the difficulty in determining $W$ after breakdown, along with other causes of sample-to-sample variation discussed in the main text. Therefore, we let $R_C W = 200$ to 2000 $\Omega$-µm (typical values for "good" and "bad" contacts respectively) for extracting the upper and lower bounds of average carrier velocity and thermal conductivity in Figure 4a,b. We use TLM to extract contact resistance for the exfoliated devices, but due to the limited amount of breakdown data, we use resistance values from low-field measurements of devices of varying length from the same sample. The average $R_C W$ for the exfoliated graphene devices in this work is ~1800 $\Omega$-µm. We vary $R_C W$ by 50% to provide bounds in Figure 4a,b similar to the case with CVD graphene. A possible reason for the exfoliated graphene having a larger contact resistance than that of the CVD graphene is the additional anneal in vacuum at 200 °C which the CVD graphene samples underwent, but the exfoliated graphene did not.

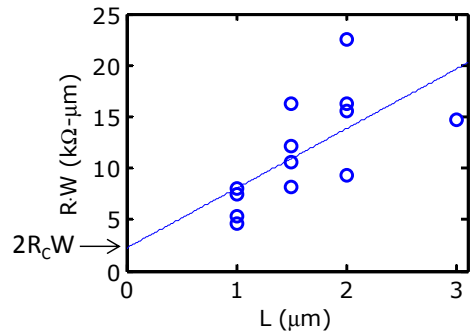

**Figure S2.** $R \cdot W$ vs. $L$ for multiple CVD graphene devices taken at the breakdown point in vacuum. The dashed line is a linear fit where the *y*-intercept corresponds to $2R_C W \approx 2200$ $\Omega$-um.

## C.  Suspended graphene device modeling

The model used to provide the simulations shown in Figure 3b is based on applying our suspended device geometry to the models developed in previous works.[4, 5] First, the gate capacitance ($C_g$) is determined by the series combination of the air gap ($t_{air} \approx 200$  nm) and the remaining SiO$_2$ ($t_{ox} \approx 100$ nm), where $C_{air} = \varepsilon_{air}\varepsilon_0/t_{air}$, $C_{ox} = \varepsilon_{ox}\varepsilon_0/t_{ox}$, and $C_g = (C_{air}^{-1} + C_{ox}^{-1})^{-1}$. We obtain $C_G \approx 4$ nF/cm$^2$ for our typical device geometry, which results in a gate induced charge $n_{cv} = C_G V_{G0}/q \le 2.5 \times 10^{11}$ cm$^{-2}$ for $V_{G0} \le 10$ V. (Here $V_{G0} = V_G - V_0$, where $V_0$ is the Dirac voltage where the sample resistance is maximum.) Second, we simulate a suspended channel region by setting the thermal conductance to the substrate to zero (i.e., $g \approx 0$), but we still allow for heat loss to the substrate underneath the contacts. Third, we adjust our model of drift velocity satura-tion in order to more accurately represent a suspended graphene sheet. Substrate effects should no longer limit transport at high fields so we assume the saturation velocity ($v_{sat}$) is determined by the zone-edge optical phonon (OP), $\hbar\omega_{OP} = 160$ meV.[6] We also use a Fermi velocity that is dependent on carrier density due to changes in the linear energy spectrum near the neutrality point for suspended graphene, $v_F(n) \sim v_0[1 + \ln(n_0/n)/4]$, where $v_0 = 0.85 \times 10^6$ m/s and $n_0 = 5 \times 10^{12}$ cm$^{-2}$ (ref 7). Fourth, the thermal generation of carriers, $n_{th} = \alpha(\pi/6)(k_B T/\hbar v_F)^2$, is given a slightly stronger than $T^2$ dependence above room temperature (although it decays back to a $T^2$ dependence at high $T$) by introducing $\alpha = 1 + e^{-(T/T_0 - 1)/2}\sqrt{T/T_0 - 1}$, where $T_0 = 300$ K. This empirical fit is used to account for the possible electron-hole pair generation from optical phonon decay[8] and/or Dirac voltage shifting that may occur during high-field device operation. Lastly, we add a contribution to the carrier density near the contacts ($n_c$) to account for the modification of the graphene electronic structure by the metal contacts ($n_c = 10^{11}$ cm$^{-2}$ at the contact but exponentially decreases away from the contact with a decay length of ~200 nm).[9]

As discussed in the main text, for the simulated curves in Figure 3b we varied the room-temperature low-field mobility ($\mu_0$) from 2,500–25,000 cm$^2$V$^{-1}$s$^{-1}$ along with the temperature dependence of the mobility ($\mu \sim T^\beta$ where $\beta$ varies from 1.5–2.5). This is meant to represent the range of "dirty" to "clean" devices that we measured experimentally. We accordingly vary the room-temperature thermal conductivity ($\kappa_0$) from 2000–3000 Wm$^{-1}$K$^{-1}$ and the breakdown temperature ($T_{BD}$) from 1420–2860 K, respectively. Consequently, the simulations show a range of breakdown current densities and electric fields comparable to those observed experimentally.

### D.  Breakdown in vacuum, air, and O$_2$ environments

In Figure S3a-c we compare the electrical breakdown of suspended graphene devices in vacuum (~10$^{-5}$ Torr), air, and O$_2$ environments. Device failure in vacuum occurs instantaneously corresponding to a sudden drop in current over a very narrow range of voltage. However, breakdown in air and O$_2$ is a more gradual process where the current degrades over a relatively wide voltage range. We expect that the very low O$_2$ partial pressure in vacuum allows for the suspended device to reach higher temperatures (> 2000 K) without oxidation degrading the device, while in air and O$_2$ oxidation may occur sporadically at lower temperatures (< 1000 K) due to the much greater availability of O$_2$. We also note that the breakdown location observed by SEM (Figure

S3d-f) is in the center of the graphene channel, corresponding to the position of maximum temperature predicted by our thermal model (see main text).

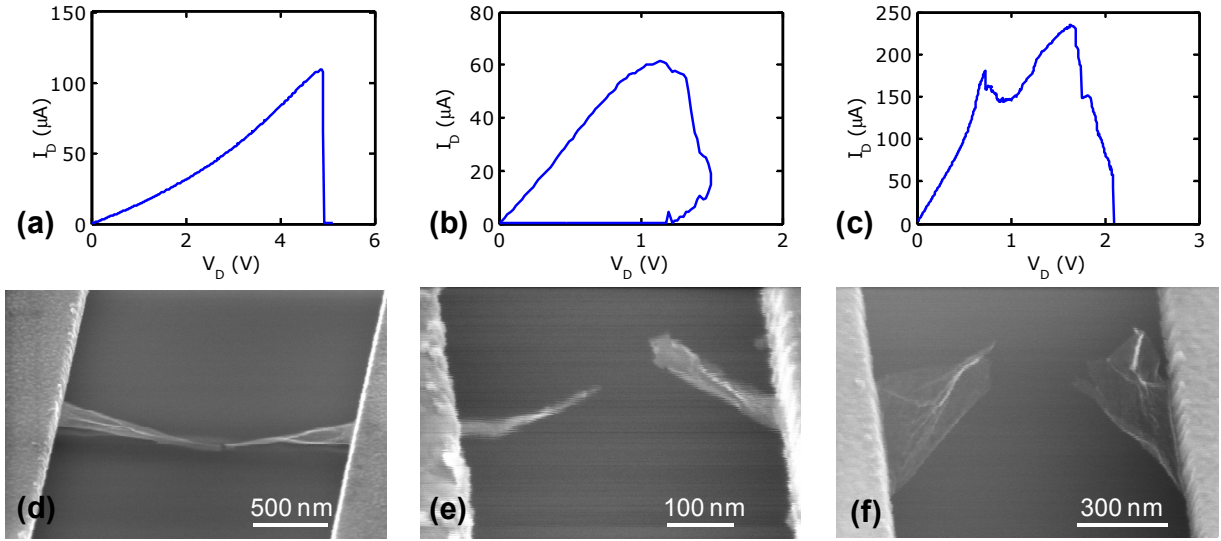

**Figure S3.** $I_D$ vs. $V_D$ and corresponding SEM images, taken at a 70° tilt with respect to the substrate, of suspended graphene broken in (a,d) vacuum, (b,e) air, and (c,f) $O_2$. Breakdown in (a) vacuum is relatively sudden and less gradual than breakdown in (b) air or (c) $O_2$.

To estimate the breakdown temperature of suspended graphene devices in vacuum, we compare breakdowns in air and vacuum. In air, solving for $T(x)$ from the heat diffusion equation (eq. 1 of the main text) results in

$$T(x) = T_0 + \frac{P}{m^2 \kappa L W t}\left(1 - \frac{\cosh(mx)}{\cosh\left(mL/2\right)}\right),\tag{S1}$$

such that the breakdown temperature is given by

$$T_{BD,air} = T_0 + \frac{P_{BD}\zeta}{m^2 \kappa L W t},\tag{S2}$$

where $m = (2g/\kappa t)^{1/2}$ and $\zeta = 1 - 1/\cosh(mL/2)$. In vacuum, heat losses due to radiation and convection are negligible[10, 11] so we use $g = 0$ which results in

$$T(x) = T_0 + \frac{PL}{8\kappa W t}\left(1 - \left(\frac{2x}{L}\right)^2\right),\tag{S3}$$

and

$$T_{BD,vac} = T_0 + \frac{P_{BD}L}{8\kappa Wt}. \tag{S4}$$

Although electrical breakdown of graphene in air is often gradual and consisting of a series of partial breaks, as shown above in Figure S3b, we can carefully choose the breakdown points from our measurements, and assuming $\kappa_{air} \approx \kappa_{vac}$, we estimate $T_{BD,vac} \approx 2230$ K.

We acknowledge some uncertainty in assuming $\kappa_{air} \approx \kappa_{vac}$ but note that the devices used for this comparison underwent similar processing. We expect graphene in vacuum to be relatively "clean", especially after current annealing, and have a higher thermal conductivity than that of graphene in air, but devices in vacuum operate at a higher average temperature, which would cause a decrease in thermal conductivity. Quantitatively evaluating these competing effects is difficult, particularly the cleanliness of a sample, thus we aim to provide upper and lower bounds for our estimate of $T_{BD,vac}$. We estimate a lower bound for $T_{BD,vac}$ by using $g = 0$ as a lower limit for the heat transfer coefficient in air,[12] and $T_{BD,air} \approx 400$ °C to account for the tendency of partial breakdown in air. We estimate an upper bound for $T_{BD,vac}$ by using $g = 10^5$ Wm$^{-2}$K$^{-1}$ in air, the theoretical upper limit based on kinetic theory,[12] and the typical $T_{BD,air} \approx 600$ °C. Thus, the extreme lower and upper bounds for $T_{BD,vac}$ are 1420 and 2860 K respectively. The lower limit appears to be a conservative estimate since the breakdown power scaled with device dimensions is typically ~3 times higher in vacuum than in air. The upper limit is comparable to the 2800 °C (i.e., 3073 K) previously estimated as the breakdown temperature of suspended graphitic nanoribbons under Joule heating.[13] Also, suspended CVD graphene has been reported to be thermally stable up to at least 2600 K.[14] However, we note these previous studies were performed in a transmission electron microscope (TEM), which is capable of obtaining lower vacuum levels than our probe station, accounting for samples likely reaching higher temperatures.

## E. Suspended graphene nanoconstriction with high on/off

We see from the SEM image after device breakdown in vacuum (Figure S3d) that the edges of the graphene channel are damaged (i.e., twisted or burned away) during the breakdown process. Burning away of the edges may also occur before breakdown during the current annealing process and result in the formation of a graphene nanoconstriction.[15] Here we show the behavior of a suspended CVD graphene device ($L = 1$ μm) with a nanoconstriction formed by current annealing. (This device was not used for extracting the data in Figure 4 of the main text.)

Figure S4a displays the measured $I_D$-$V_G$ of a nanoconstriction device at $T = 80$–300 K for $V_D$ = 200 mV, showing high on/off > $10^3$ at room temperature and > $10^9$ at $T = 80$ K. In Figure S4b we vary the drain bias at $T = 150$ K to show that the effective band gap and on/off is diminished at high fields. We observe high on/off > $10^6$ for $V_D \leq 200$ mV, but a low on/off < 10 when we increase the bias to $V_D = 1$ V. At low bias and low temperature we observe discrete conductance peaks (e.g., $V_D = 50$ mV and $T = 150$ K in Figure S4b). It has been suggested that the regions of the graphene channel that connect the narrow nanoconstriction to the wider graphene sheet may actually be confined longitudinally (i.e., along the length of the channel) and behave as quantum dots in series.[15] Thus, the conductance peaks correspond to resonant tunneling through the quantized energy levels of these quantum dots. In Figure S4c we assume thermal activation, $I_{min} \sim$ exp(-$E_g/2k_BT$), to extract an effective band gap of $E_g \sim 0.35$ eV and corresponding width of ~12 nm, where $W = 2\pi\hbar v_F/E_g$.[16] This width extraction may be an underestimate since the aforementioned quantum dot regions may increase the effective band gap. Unfortunately the device broke before we were able to image the channel and measure the width of the nanoconstriction.

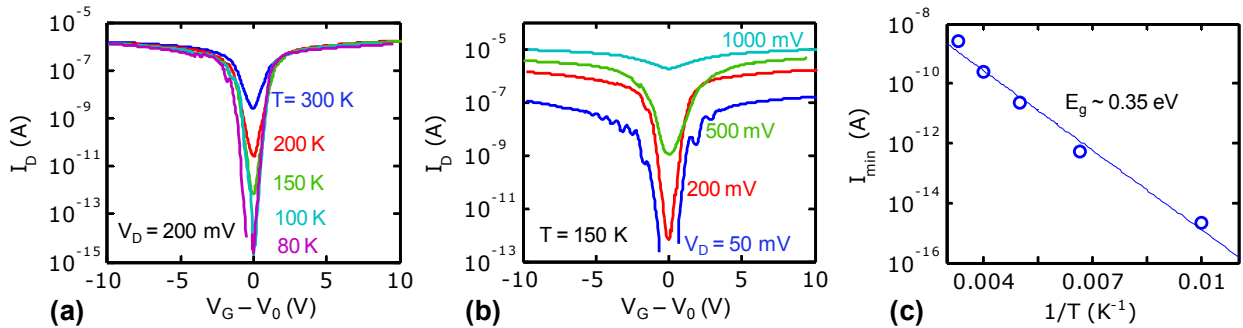

**Figure S4.** (a) Measured current vs. gate voltage at $T = 80$–300 K of a suspended CVD graphene nanoconstriction formed by current annealing. (b) Electrical transport measurements at $T = 150$ K under varying bias $V_D = 50$–1000 mV. High on/off > $10^6$ is observed for $V_D \leq 200$ mV, while an increase in bias to $V_D = 1000$ mV results in on/off < 10. (c) Temperature dependence of the minimum current at $V_D = 200$ mV. An effective band gap $E_g \sim 0.35$ eV is extracted assuming thermal activation, $I_{min} \sim$ exp(-$E_g/2k_BT$).

## Supplementary References:

10.   The power loss due to radiation is estimated by $P_{rad} = \sigma \varepsilon A (T^4 - T_0^4)$, where $\sigma = 5.67 \times 10^{-8}$ Wm$^{-2}$K$^{-4}$ is the Stefan-Boltzmann constant, $\varepsilon \sim 2.3$ % is the emissivity of graphene and assumed equal to the absorption from ref 11, $A = 2$ μm$^2$ is an upper limit for the area of the graphene channel, and $T = 1300$ K is an upper limit for average temperature along the channel. We estimate $P_{rad} \approx 7.4$ nW, which is several orders of magnitude less than the electric power that is dissipated.

# <u>Supplementary Information</u>

## High-Field Electrical and Thermal Transport in Suspended Graphene

Vincent E. Dorgan,[1,2] Ashkan Behnam,[1,2] Hiram Conley,[3] Kirill I. Bolotin,[3] and Eric Pop[1,2,4]

[1]*Micro and Nanotechnology Lab, Univ. Illinois at Urbana-Champaign, IL 61801, USA*

[2]*Dept. Electrical & Computer Eng., Univ. Illinois at Urbana-Champaign, IL 61801, USA*

[3]*Dept. Physics & Astronomy, Vanderbilt University, Nashville, TN 37235, USA*

[4]*Beckman Institute, Univ. Illinois at Urbana-Champaign, IL 61801, USA*

## A.  Graphene device fabrication and suspension

We fabricated suspended devices by two methods, one by mechanically exfoliating graphene from natural graphite, the other by chemical vapor deposition (CVD) growth on Cu substrates. With the standard "tape method," graphene is mechanically exfoliated onto a substrate of ~300 nm of $SiO_2$ with a highly doped Si substrate (*p*-type, $5 \times 10^{-3}$ Ω-cm). The tape residue is then cleaned off by annealing at 400 °C for 120 min with a flow of Ar/H (500/500 sccm) at atmospheric pressure. Monolayer graphene flakes are then identified with an optical microscope and confirmed via Raman spectroscopy (Figure S1a).[1]

Graphene growth by CVD is performed by flowing $CH_4$ and Ar gases at 1000 °C and 0.5 Torr chamber pressure, which results primarily in monolayer graphene growth on both sides of the Cu foil.[2] One graphene side is protected with a ~250 nm thick layer of polymethyl methacrylate (PMMA) while the other is removed with a 20 sccm $O_2$ plasma reactive ion etch (RIE) for 20 seconds. The Cu foil is then etched overnight in aqueous $FeCl_3$, leaving the graphene supported by the PMMA floating on the surface of the solution. The PMMA + graphene bilayer film is transferred via a glass slide to a HCl bath and then to two separate deionized water baths. Next, the film is transferred to the $SiO_2$ (~300 nm) on Si substrate (*p*-type, $5 \times 10^{-3}$ Ω-cm) and left for a few hours to dry. The PMMA is removed using a 1:1 mixture of methylene chloride and methanol, followed by a one hour Ar/$H_2$ anneal at 400 °C to remove PMMA and other organic residue.

The following fabrication steps are performed for both the exfoliated graphene and CVD graphene devices. We pattern a rectangular graphene channel using e-beam lithography and an $O_2$ plasma etch. Another e-beam lithography step is used to define the electrodes, which consist of 0.5-3 nm of Cr and 80 nm of Au. The sample is annealed again in Ar/H at 400 °C in order to help remove the polymer residue leftover from the fabrication process.

The suspension of the graphene sheet is accomplished by etching away ~200 nm of the underlying $SiO_2$.[3] The sample is placed in 50:1 BOE for 18 min followed by a deionized (DI) water bath for 5 min. Isopropyl alcohol (IPA) is squirted into the water bath while the water is poured out so that the sample always remains in liquid. After all the water has been poured out and only IPA remains, the sample is put into a critical point dryer (CPD). Following the CPD process we confirm suspension using SEM (Figures 1b-d) or AFM (Figures S1b,c). We note that the CVD graphene samples underwent a vacuum anneal at 200 °C after the suspension process. Typically, device performance improved after this anneal but we also noticed several devices would break. Due to the fragility and relatively limited number of exfoliated graphene devices, we did not perform this additional annealing step with the exfoliated graphene samples.

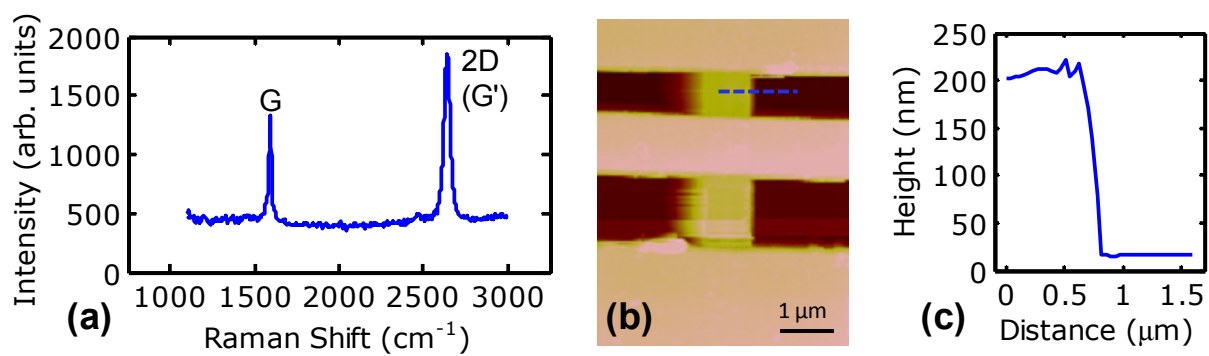

**Figure S1.** (a) Raman spectrum showing the G and 2D (also called G') bands of monolayer graphene. (b) AFM image of suspended exfoliated graphene where the dashed blue line corresponds to the (c) height vs. distance trace. The initial $SiO_2$ thickness is 300 nm, of which approximately 200 nm is etched during the suspension process.

## B.  Electrical contact resistance

Contact resistance of the CVD graphene devices is determined using the transfer length method (TLM). In Figure S2 we plot $R \cdot W$ versus $L$ and extract $R_C W \approx 1100$ Ω-um from the linear fit (dashed line). We use the resistance value at breakdown for extracting contact resistance. The spread in $R \cdot W$ values that results in a poor linear fit in Figure S2 may be associated with the difficulty in determining $W$ after breakdown, along with other causes of sample-to-sample variation discussed in the main text. Therefore, we let $R_C W = 200$ to $2000$ $\Omega$-µm (typical values for "good" and "bad" contacts respectively) for extracting the upper and lower bounds of average carrier velocity and thermal conductivity in Figure 4a,b. We use TLM to extract contact resistance for the exfoliated devices, but due to the limited amount of breakdown data, we use resistance values from low-field measurements of devices of varying length from the same sample. The average $R_C W$ for the exfoliated graphene devices in this work is ~1800 $\Omega$-µm. We vary $R_C W$ by 50% to provide bounds in Figure 4a,b similar to the case with CVD graphene. A possible reason for the exfoliated graphene having a larger contact resistance than that of the CVD graphene is the additional anneal in vacuum at 200 °C which the CVD graphene samples underwent, but the exfoliated graphene did not.

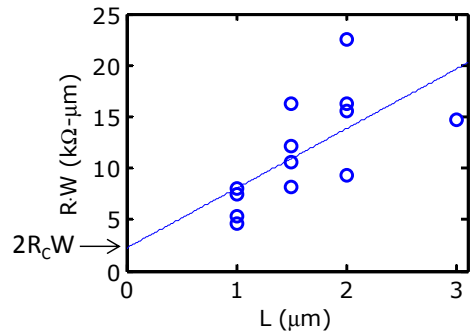

**Figure S2.** $R \cdot W$ vs. $L$ for multiple CVD graphene devices taken at the breakdown point in vacuum. The dashed line is a linear fit where the $y$-intercept corresponds to $2R_C W \approx 2200$ $\Omega$-um.

## C. Suspended graphene device modeling

The model used to provide the simulations shown in Figure 3b is based on applying our suspended device geometry to the models developed in previous works.[4, 5] First, the gate capacitance ($C_g$) is determined by the series combination of the air gap ($t_{air} \approx 200$ nm) and the remaining SiO$_2$ ($t_{ox} \approx 100$ nm), where $C_{air} = \varepsilon_{air}\varepsilon_0/t_{air}$, $C_{ox} = \varepsilon_{ox}\varepsilon_0/t_{ox}$, and $C_g = (C_{air}^{-1} + C_{ox}^{-1})^{-1}$. We obtain $C_G \approx 4$ nF/cm$^2$ for our typical device geometry, which results in a gate induced charge $n_{cv} = C_G V_{G0}/q \leq 2.5 \times 10^{11}$ cm$^{-2}$ for $V_{G0} \leq 10$ V. (Here $V_{G0} = V_G - V_0$, where $V_0$ is the Dirac voltage where the sample resistance is maximum.) Second, we simulate a suspended channel region by setting the thermal conductance to the substrate to zero (i.e., $g \approx 0$), but we still allow for heat loss to the substrate underneath the contacts. Third, we adjust our model of drift velocity satura-tion in order to more accurately represent a suspended graphene sheet. Substrate effects should no longer limit transport at high fields so we assume the saturation velocity ($v_{sat}$) is determined by the zone-edge optical phonon (OP), $\hbar\omega_{OP} = 160$ meV.[6] We also use a Fermi velocity that is dependent on carrier density due to changes in the linear energy spectrum near the neutrality point for suspended graphene, $v_F(n) \sim v_0[1 + \ln(n_0/n)/4]$, where $v_0 = 0.85 \times 10^6$ m/s and $n_0 = 5 \times 10^{12}$ cm$^{-2}$ (ref 7). Fourth, the thermal generation of carriers, $n_{th} = \alpha(\pi/6)(k_B T/\hbar v_F)^2$, is given a slightly stronger than $T^2$ dependence above room temperature (although it decays back to a $T^2$ dependence at high $T$) by introducing $\alpha = 1 + e^{-(T/T_0-1)/2}\sqrt{T/T_0 - 1}$, where $T_0 = 300$ K. This empirical fit is used to account for the possible electron-hole pair generation from optical phonon decay[8] and/or Dirac voltage shifting that may occur during high-field device operation. Lastly, we add a contribution to the carrier density near the contacts ($n_c$) to account for the modification of the graphene electronic structure by the metal contacts ($n_c = 10^{11}$ cm$^{-2}$ at the contact but exponentially decreases away from the contact with a decay length of ~200 nm).[9]

As discussed in the main text, for the simulated curves in Figure 3b we varied the room-temperature low-field mobility ($\mu_0$) from 2,500–25,000 cm$^2$V$^{-1}$s$^{-1}$ along with the temperature dependence of the mobility ($\mu \sim T^\beta$ where $\beta$ varies from 1.5–2.5). This is meant to represent the range of "dirty" to "clean" devices that we measured experimentally. We accordingly vary the room-temperature thermal conductivity ($\kappa_0$) from 2000–3000 Wm$^{-1}$K$^{-1}$ and the breakdown temperature ($T_{BD}$) from 1420–2860 K, respectively. Consequently, the simulations show a range of breakdown current densities and electric fields comparable to those observed experimentally.

## D. Breakdown in vacuum, air, and O$_2$ environments

In Figure S3a-c we compare the electrical breakdown of suspended graphene devices in vacuum (~10$^{-5}$ Torr), air, and O$_2$ environments. Device failure in vacuum occurs instantaneously corresponding to a sudden drop in current over a very narrow range of voltage. However, breakdown in air and O$_2$ is a more gradual process where the current degrades over a relatively wide voltage range. We expect that the very low O$_2$ partial pressure in vacuum allows for the suspended device to reach higher temperatures (> 2000 K) without oxidation degrading the device, while in air and O$_2$ oxidation may occur sporadically at lower temperatures (< 1000 K) due to the much greater availability of O$_2$. We also note that the breakdown location observed by SEM (Figure

S3d-f) is in the center of the graphene channel, corresponding to the position of maximum temperature predicted by our thermal model (see main text).

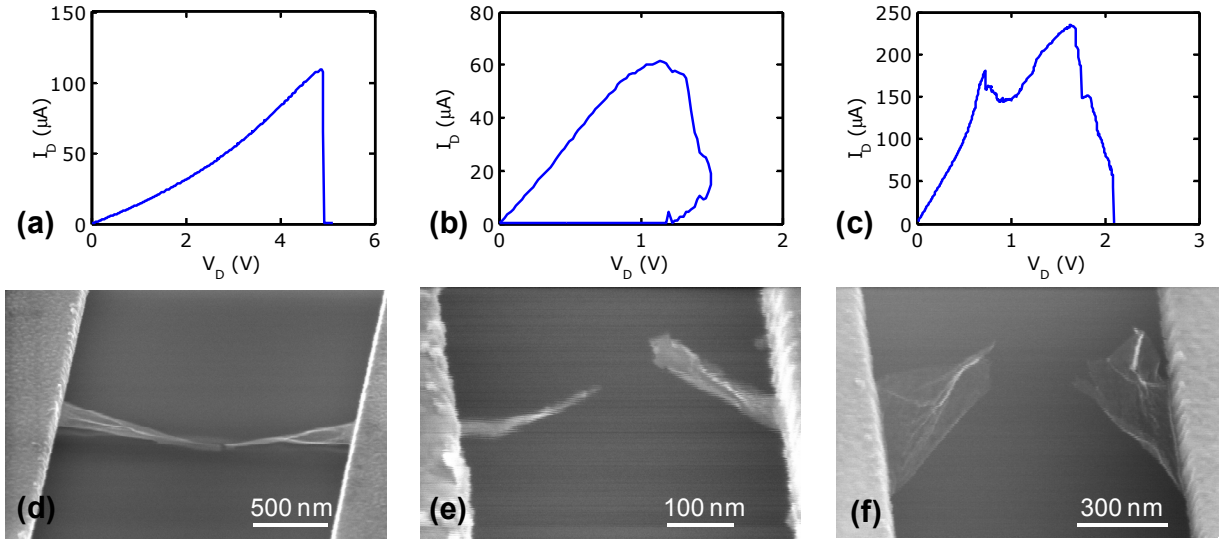

**Figure S3.** $I_D$ vs. $V_D$ and corresponding SEM images, taken at a 70° tilt with respect to the substrate, of suspended graphene broken in (a,d) vacuum, (b,e) air, and (c,f) $O_2$. Breakdown in (a) vacuum is relatively sudden and less gradual than breakdown in (b) air or (c) $O_2$.

To estimate the breakdown temperature of suspended graphene devices in vacuum, we compare breakdowns in air and vacuum. In air, solving for $T(x)$ from the heat diffusion equation (eq. 1 of the main text) results in

$$T(x) = T_0 + \frac{P}{m^2 \kappa L W t}\left(1 - \frac{\cosh(mx)}{\cosh(mL/2)}\right), \tag{S1}$$

such that the breakdown temperature is given by

$$T_{BD,air} = T_0 + \frac{P_{BD}\,\zeta}{m^2 \kappa L W t}, \tag{S2}$$

where $m = (2g/\kappa t)^{1/2}$ and $\zeta = 1 - 1/\cosh(mL/2)$. In vacuum, heat losses due to radiation and convection are negligible[10, 11] so we use $g = 0$ which results in

$$T(x) = T_0 + \frac{PL}{8\kappa W t}\left(1 - \left(\frac{2x}{L}\right)^2\right), \tag{S3}$$

and

$$T_{BD,vac} = T_0 + \frac{P_{BD}L}{8\kappa Wt}.$$  (S4)

Although electrical breakdown of graphene in air is often gradual and consisting of a series of partial breaks, as shown above in Figure S3b, we can carefully choose the breakdown points from our measurements, and assuming $\kappa_{air} \approx \kappa_{vac}$, we estimate $T_{BD,vac} \approx 2230$ K.

We acknowledge some uncertainty in assuming $\kappa_{air} \approx \kappa_{vac}$ but note that the devices used for this comparison underwent similar processing. We expect graphene in vacuum to be relatively "clean", especially after current annealing, and have a higher thermal conductivity than that of graphene in air, but devices in vacuum operate at a higher average temperature, which would cause a decrease in thermal conductivity. Quantitatively evaluating these competing effects is difficult, particularly the cleanliness of a sample, thus we aim to provide upper and lower bounds for our estimate of $T_{BD,vac}$. We estimate a lower bound for $T_{BD,vac}$ by using $g = 0$ as a lower limit for the heat transfer coefficient in air,[12] and $T_{BD,air} \approx 400$ °C to account for the tendency of partial breakdown in air. We estimate an upper bound for $T_{BD,vac}$ by using $g = 10^5$ Wm$^{-2}$K$^{-1}$ in air, the theoretical upper limit based on kinetic theory,[12] and the typical $T_{BD,air} \approx 600$ °C. Thus, the extreme lower and upper bounds for $T_{BD,vac}$ are 1420 and 2860 K respectively. The lower limit appears to be a conservative estimate since the breakdown power scaled with device dimensions is typically ~3 times higher in vacuum than in air. The upper limit is comparable to the 2800 °C (i.e., 3073 K) previously estimated as the breakdown temperature of suspended graphitic nanoribbons under Joule heating.[13] Also, suspended CVD graphene has been reported to be thermally stable up to at least 2600 K.[14] However, we note these previous studies were performed in a transmission electron microscope (TEM), which is capable of obtaining lower vacuum levels than our probe station, accounting for samples likely reaching higher temperatures.

## E.  Suspended graphene nanoconstriction with high on/off

We see from the SEM image after device breakdown in vacuum (Figure S3d) that the edges of the graphene channel are damaged (i.e., twisted or burned away) during the breakdown process. Burning away of the edges may also occur before breakdown during the current annealing process and result in the formation of a graphene nanoconstriction.[15] Here we show the behavior of a suspended CVD graphene device ($L = 1$ μm) with a nanoconstriction formed by current annealing. (This device was not used for extracting the data in Figure 4 of the main text.)

Figure S4a displays the measured $I_D$-$V_G$ of a nanoconstriction device at $T = 80$–$300$ K for $V_D$ = 200 mV, showing high on/off > $10^3$ at room temperature and > $10^9$ at $T = 80$ K. In Figure S4b we vary the drain bias at $T = 150$ K to show that the effective band gap and on/off is diminished at high fields. We observe high on/off > $10^6$ for $V_D \leq 200$ mV, but a low on/off < 10 when we increase the bias to $V_D = 1$ V. At low bias and low temperature we observe discrete conductance peaks (e.g., $V_D = 50$ mV and $T = 150$ K in Figure S4b). It has been suggested that the regions of the graphene channel that connect the narrow nanoconstriction to the wider graphene sheet may actually be confined longitudinally (i.e., along the length of the channel) and behave as quantum dots in series.[15] Thus, the conductance peaks correspond to resonant tunneling through the quantized energy levels of these quantum dots. In Figure S4c we assume thermal activation, $I_{min} \sim$ exp(-$E_g/2k_BT$), to extract an effective band gap of $E_g \sim 0.35$ eV and corresponding width of ~12 nm, where $W = 2\pi\hbar v_F/E_g$.[16] This width extraction may be an underestimate since the aforementioned quantum dot regions may increase the effective band gap. Unfortunately the device broke before we were able to image the channel and measure the width of the nanoconstriction.

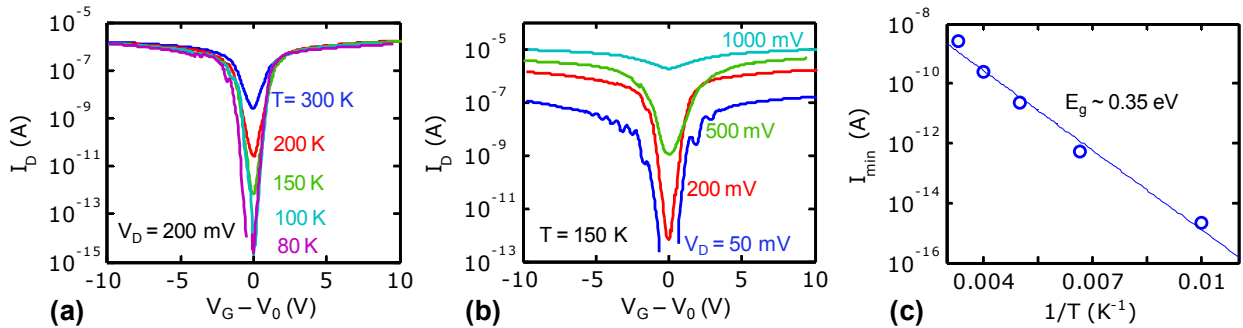

**Figure S4.** (a) Measured current vs. gate voltage at $T = 80$–$300$ K of a suspended CVD graphene nanoconstriction formed by current annealing. (b) Electrical transport measurements at $T = 150$ K under varying bias $V_D = 50$–$1000$ mV. High on/off > $10^6$ is observed for $V_D \leq 200$ mV, while an increase in bias to $V_D = 1000$ mV results in on/off < 10. (c) Temperature dependence of the minimum current at $V_D = 200$ mV. An effective band gap $E_g \sim 0.35$ eV is extracted assuming thermal activation, $I_{min} \sim$ exp(-$E_g/2k_BT$).

## Supplementary References:

10.   The power loss due to radiation is estimated by $P_{rad} = \sigma \varepsilon A (T^4 - T_0^4)$, where $\sigma = 5.67 \times 10^{-8}$ $Wm^{-2}K^{-4}$ is the Stefan-Boltzmann constant, $\varepsilon \sim 2.3$ % is the emissivity of graphene and assumed equal to the absorption from ref 11, $A = 2$ $\mu m^2$ is an upper limit for the area of the graphene channel, and $T = 1300$ K is an upper limit for average temperature along the channel. We estimate $P_{rad} \approx 7.4$ nW, which is several orders of magnitude less than the electric power that is dissipated.